\documentclass[lettersize,journal]{IEEEtran}
\usepackage{amsmath,amsfonts}
\usepackage{algorithm}
\usepackage{array}
\usepackage[caption=false,font=normalsize,labelfont=sf,textfont=sf]{subfig}
\usepackage{textcomp}
\usepackage{stfloats}
\usepackage{url}
\usepackage{verbatim}
\usepackage{graphicx}
\usepackage{cite}
\usepackage{makecell}

\usepackage{amsthm,amsmath,amssymb}
\usepackage{makecell}
\usepackage{mathrsfs}
  \usepackage{float}
  \usepackage{algorithm}
\usepackage{algpseudocode}
\usepackage{amsmath}
\usepackage{bm}
\usepackage{color}
\usepackage{comment}
\usepackage{booktabs}
\usepackage{leftidx}
\usepackage{epstopdf}
\usepackage{amsmath}
\usepackage{makecell}
\usepackage{multirow}
\usepackage{hyperref}
\usepackage{amssymb}
\usepackage{adjustbox}
\usepackage{booktabs}
\usepackage{tabularx}
\usepackage{ragged2e}
\newcolumntype{L}[1]{>{\RaggedRight\arraybackslash}p{#1}}
\newcolumntype{Y}{>{\RaggedRight\arraybackslash}X}

\begin{document}

\title{What Makes Good Datasets for Intelligent Multi-Modal Sensing-Communication Integration?}

\author{Mingran Sun,~\IEEEmembership{Graduate Student Member,~IEEE}, Zengrui Han,~\IEEEmembership{Graduate Student Member,~IEEE}, Xuanyu Liu,~\IEEEmembership{Graduate Student Member,~IEEE}, Xuesong Cai,~\IEEEmembership{Senior Member,~IEEE}, and Xiang Cheng,~\IEEEmembership{Fellow,~IEEE}

\thanks{M. Sun, Z. Han, X. Liu, X. Cai, and X. Cheng are with the State Key Laboratory of Photonics and Communications, School of Electronics, Peking University, Beijing 100871, China (e-mail: mingransun@stu.pku.edu.cn; zengruihan@stu.pku.edu.cn; xyliu25@stu.pku.edu.cn; xuesong.cai@pku.edu.cn; xiangcheng@pku.edu.cn).}

}



\maketitle 

\begin{abstract}
Intelligent multi-modal sensing-communication integration, namely Synesthesia of Machines (SoM), represents an AI-native paradigm for enabling embodied intelligence and future wireless networks. Multi-modal sensing and communication foundation models developed for SoM learn cross-modal associations and generalizable environmental representations through large-scale pretraining, making the quality of pretraining datasets critical. However, existing methods lack unified evaluation of heterogeneous multi-modal data and thus struggle to meet the requirements for multi-modal pretraining dataset construction and selection. To bridge this gap, we introduce a comprehensive data evaluation framework for intelligent multi-modal sensing-communication integration. Specifically, we establish two complementary heterogeneous multi-modal data evaluation dimensions, including intrinsic data quality and validity, and model-based data utility. Furthermore, we present case studies to validate the effectiveness of the two evaluation dimensions. We also extend our analysis to investigate the impact of data quality on pretraining scaling laws and explore the value of integrating measurement and simulation data. Finally, we outline promising directions for future research on data evaluation.
\end{abstract}

\section{Introduction}
\IEEEPARstart{A}{s} embodied intelligence moves into the physical world, autonomous vehicles, uncrewed aerial vehicles (UAVs), robots, and wearable devices face growing demands for environmental awareness and connectivity \cite{tase, jcise}. Sensing and communication are fundamental to how these systems perceive their surroundings, connect with others, and act \cite{wcm-agi, CAAI}. However, current research on embodied systems relies primarily on vision for environmental understanding and treats communications chiefly as a means of data transmission. Developed largely in isolation, these two capabilities fall short of supporting reliable interaction with complex environments. To fully exploit the complementary views of the physical world offered by different frequency bands and modalities, synesthesia of machines (SoM) has been proposed \cite{som}. SoM refers to intelligent multi-modal sensing-communication integration, representing an AI-native paradigm for enabling embodied intelligence and future wireless networks. By jointly exploiting complementary information across electromagnetic frequency bands, it enables sensing and communication to reinforce each other, bringing previously separate perception, connectivity, and intelligent decision-making together in a cooperative system. First, SoM enables \textbf{a shift from single frequency band observations to complementary observations across the electromagnetic spectrum}, overcoming the limitations of any single modality in the face of occlusion, range, resolution, and changing environmental conditions. Second, SoM enables \textbf{a shift from emulating to surpassing human perception}. RF signals, radar, and infrared sensing provide information beyond direct human perception, revealing motion, signal propagation, thermal radiation, and non-line-of-sight environments. Third, SoM enables \textbf{a shift from individual intelligence to networked embodied intelligence}. Devices can share observations across locations and viewpoints to build a collective understanding of their environment, which in turn helps optimize connectivity and transmission.

Realizing SoM requires learning a shared representation of the physical environment and the relationships among modalities from heterogeneous electromagnetic observations. To this end, multi-modal sensing and communication foundation models trained at scale to capture generalizable environmental representations across electromagnetic frequency bands offer a promising path toward intelligent multi-modal sensing-communication integration \cite{som-tnse}. The benefits of multi-modal pretraining extend beyond joint multi-modal input. By learning cross-modal associations during training, the model can form a richer understanding of the environment from a single modality at deployment. This transfers the gains of multi-modal learning to uni-modal inference without the computation or latency of processing additional modalities. The capabilities of the multi-modal sensing and communication foundation model depend critically on its pretraining data. Such a model requires large-scale data spanning vision, light detection and ranging (LiDAR), radar, and channel state information (CSI), as well as diverse scenarios, devices, and propagation conditions. Nonetheless, data volume alone does not determine data value. Physical fidelity, cross-modal consistency, scenario coverage, and information redundancy all shape the environmental knowledge and cross-modal associations the model can learn. Assessing the quality of multi-modal sensing and communication data is essential to dataset construction and generation, pretraining data selection, and the development of more capable foundation models. Therefore, a comprehensive framework for evaluating multi-modal sensing and communication data is urgently needed.

In recent years, researchers have investigated data evaluation for foundation models across a range of domains. In natural language processing (NLP) domain, data evaluation expanded from intrinsic properties, such as correctness, completeness, and diversity, to model utility measured through training outcomes, with evaluation results increasingly used to guide data filtering and selection. In computer vision (CV) and multi-modal perception domain, research addressed image and annotation quality, scenario coverage, and the synchronization, calibration, and cross-modal correspondence of camera, LiDAR, and radar data, particularly in autonomous driving. Recent wireless research examined similarity among CSI datasets using distributional distances, reduced-dimensional representations, and pretrained model representations, with the aim of predicting cross-dataset transfer performance. Nevertheless, extending the methods in NLP, CV and multi-modal perception, and wireless communication domains to evaluating multi-modal sensing and communication data reveals two following limitations.

\begin{itemize}
    \item \textbf{Insufficient intrinsic quality evaluation of heterogeneous multi-modal data}: Existing methods mainly focus on single-modality quality or sensor synchronization and calibration, lacking systematic evaluation of radio-frequency (RF) channel fidelity, cross-modal spatiotemporal consistency, physical coherence, and information complementarity. The absence of explicit environmental representations in RF channel data further complicates unified multi-modal quality assessment.
    \item \textbf{Insufficient model-based and task-oriented data utility evaluation}: The diversity and complexity of embodied sensing-communication models and tasks make data utility evaluation increasingly challenging. Existing methods struggle to evaluates the contributions of different datasets to model learning, downstream sensing-communication task performance, and multi-modal sensing and communication foundation model pretraining.s
\end{itemize}

To fill these gaps, we propose a comprehensive framework of evaluating data for multi-modal sensing and communication foundation models. In Section II, we first review data evaluation methods in NLP, CV and multi-modal perception, and wireless communication domains, highlighting the challenges and opportunities in multi-modal sensing and communication data evaluation. In Section III and Section IV, we then present the framework along two dimensions, i.e., the intrinsic quality and validity, and model-based data utility. In Section V, we further illustrate the framework through a controlled case study to validate the effectiveness of the proposed framework in evaluating multi-modal sensing and communication data. Finally, in Section VI and Section VII, we conclude and outline promising directions for future research.

\section{Cross-Domain Perspectives on Data Evaluation for Foundation Models}
This section reviews data evaluation in NLP, CV and multi-modal perception, and wireless communications, as shown in Table~\ref{tab:cross_domain_data_evaluation}. It compares their intrinsic and model-based criteria and identifies the requirements for a physically grounded evaluation framework for multi-modal sensing and communication foundation models.

\begin{table*}[t]
\centering
\caption{Comparison of data evaluation across research domains}
\label{tab:cross_domain_data_evaluation}
\footnotesize
\setlength{\tabcolsep}{4pt}
\renewcommand{\arraystretch}{1.22}
\begin{tabularx}{\textwidth}{@{}
    L{1.65cm}
    L{1.65cm}
    L{2.05cm}
    L{2.35cm}
    Y
    Y
    @{}}
\toprule
\textbf{Domain} &
\textbf{Modalities} &
\textbf{Data sources} &
\textbf{Conditions} &
\textbf{Evaluation criteria} &
\textbf{Assessment} \\
\midrule

NLP~\cite{chim2025synthetic,iskander2024quality} &
Text &
User-generated and synthetic text &
Topics, instructions, and tasks &
Correctness, diversity, learning value, and downstream performance &
Links intrinsic quality to model utility. Does not assess physical validity or sensor alignment. \\

\addlinespace[3pt]
Vision and multi-modal perception~\cite{caesar2020nuscenes,gadre2023datacomp} &
Image, text, LiDAR, and radar &
Web data and measured sensor data &
Scenarios, viewpoints, and sensor setups &
Annotation quality, synchronization, calibration, cross-modal correspondence, and downstream performance &
Accounts for sensor consistency and controlled training outcomes. Does not evaluate wireless propagation or communication tasks. \\

\addlinespace[3pt]
Wireless communications~\cite{wirelesssimilarity2026,lwmcde2026} &
Primarily CSI &
Simulated and measured wireless data &
Propagation environments and radio configurations &
Dataset distance, representation similarity, and cross-dataset transfer &
Supports transfer prediction and data selection. Largely limited to CSI and lacks systematic treatment of source and size heterogeneity. \\

\addlinespace[3pt]
\textbf{Proposed framework} &
Image, map, LiDAR, radar, and CSI &
Measured, simulated, and generated data &
Scenarios, bands, devices, system configurations, and modality combinations &
Physical validity, alignment, coverage, redundancy, model utility, transfer, robustness, and data efficiency&
Combines intrinsic and model-based evaluation across heterogeneous multi-modal data. Requires consistent metadata and controlled training comparisons. \\

\bottomrule
\end{tabularx}
\end{table*}

\subsection{Data Evaluation in Natural Language Processing}
NLP domain established the basic rationale for intrinsic data evaluation. Data are assessed through properties observable without downstream training, including grammatical well-formedness, factual or procedural correctness, completeness, fluency, duplication, and topical diversity. For synthetic text, these criteria determine whether generated content is coherent, faithful to its intended source or instruction, and sufficiently varied to represent the target distribution. Intrinsic evaluation therefore treats the data record itself as the object of assessment and provides a principled basis for detecting invalid, repetitive, or otherwise defective examples before corpus construction.

Subsequent work extended this intrinsic perspective toward extrinsic, or model-based, evaluation. The question becomes whether training on an example produces measurable improvement in the target model. Studies of synthetic text explicitly distinguish intrinsic criteria from extrinsic outcomes~\cite{chim2025synthetic}. For tool-using language models, candidate examples are first screened for correctness and learning value and are then assessed through controlled training~\cite{iskander2024quality}. With architecture, optimization, data volume, and compute fixed, convergence, held-out performance, reasoning, and tool-use success provide evidence of utility. The resulting feedback can guide filtering, mixture design, and further collection, thereby connecting data properties with the capabilities that the model actually acquires.

The resulting NLP methodology cannot be transferred directly to sensing and communication data. Language data are judged mainly by semantic and task-level criteria, whereas wireless observations are constrained by propagation laws, antenna geometry, synchronization, carrier frequency, bandwidth, and system configuration. A sample may appear statistically plausible while violating a physical constraint or representing a different wireless channel. Multi-modal sensing and communication data evaluation must therefore combine model utility with domain-specific validity tests.

\subsection{Data Evaluation in Computer Vision and Multi-Model Perception}
In CV and multi-modal perception, intrinsic evaluation has progressed from image appearance, annotation accuracy, and scene diversity to the physical consistency of multiple sensors. nuScenes jointly collects camera, LiDAR, and radar measurements with explicit calibration and temporal synchronization~\cite{caesar2020nuscenes}. Cross-modal agreement consequently requires more than semantic correspondence. Observations should describe the same objects, locations, and dynamic state while preserving their spatial and temporal relationships. This requirement also applies when visual observations are paired with RF measurements.

The field has further established model-based evaluation under controlled training conditions. DataComp fixes the model family, training recipe, and computational scale, then compares filtering and dataset-construction strategies across downstream tests~\cite{gadre2023datacomp}. Classification, detection, retrieval, distribution-shift robustness, and data efficiency quantify each corpus's practical contribution. The protocol shows that image quality and dataset size alone are insufficient. Intrinsic properties and extrinsic utility are complementary dimensions.

The limitation for data evaluation in CV and multi-modal perception domain lies in the task domain and physical model of existing benchmarks. Their evaluations emphasize visual recognition, synchronization, calibration, and observation consistency, but omit electromagnetic propagation, transmitter-receiver geometry, radio-system configuration, and communication-sensing coupling. Improvement in visual detection therefore does not establish value for channel estimation, beam selection, localization, or other joint communication-perception tasks.

\subsection{Data Evaluation in Wireless Communications}
In wireless communications, recent work has begun to characterize data through dataset similarity. One direction compares CSI distribution distances using UMAP-based representations to estimate transfer between environments~\cite{wirelesssimilarity2026}. Another measures similarity in a representation space induced by a pre-trained wireless language model and improves the metric through contrastive learning~\cite{lwmcde2026}. These studies suggest that task-relevant representations can support data selection before full-scale pretraining.

Representation-based similarity can therefore support pretraining data selection, redundancy analysis, and identification of distributional gaps. Distances in a learned wireless space may indicate transferability and reveal whether an additional CSI collection contributes conditions absent from an existing corpus. This provides a low-cost screening stage before end-to-end training, although interpretation remains dependent on the representation model's scope and inductive bias.

The principal limitation is that these methods remain single-modal. They do not jointly evaluate CSI with vision, LiDAR, or radar, and they do not explicitly account for heterogeneous sources or dataset sizes. Therefore, it cannot support the dataset selection during the pretraining of multi-modal sensing and communication foundation models.

\subsection{Challenges and Opportunities}
Overall, two challenges remain central. First, multi-modal sensing and communication data evaluation must be physically grounded. A suitable framework should assess the integrity of each modality together with temporal synchronization, spatial alignment, semantic consistency, and electromagnetic consistency across vision, LiDAR, radar, and CSI. In particular, it should determine whether apparently corresponding observations are generated by the same physical environment, state, and propagation condition. Second, the value of heterogeneous data must be compared fairly across measurement, simulation, and generation sources, as well as across different devices, frequencies, scenes, and dataset sizes. The evaluation must distinguish genuinely informative additions from repetition, source artifacts, and data that are irrelevant to the intended deployment domain.

These considerations motivate a comprehensive evaluation framework comprising two coupled dimensions. The first dimension concerns intrinsic data properties and determines whether samples are complete, physically valid, mutually consistent, and informative relative to the existing corpus. The second concerns extrinsic model utility and measures contributions to multi-task capability, cross-domain transfer, robustness, data efficiency, and system-level performance under controlled training conditions. Together, the two dimensions provide a principled basis for constructing, filtering, and expanding pretraining corpora for multi-modal sensing and communication foundation models.

\section{Dimension 1: Intrinsic Data Quality and Validity}

\begin{figure*}
\centerline{\includegraphics[width=0.9\textwidth]{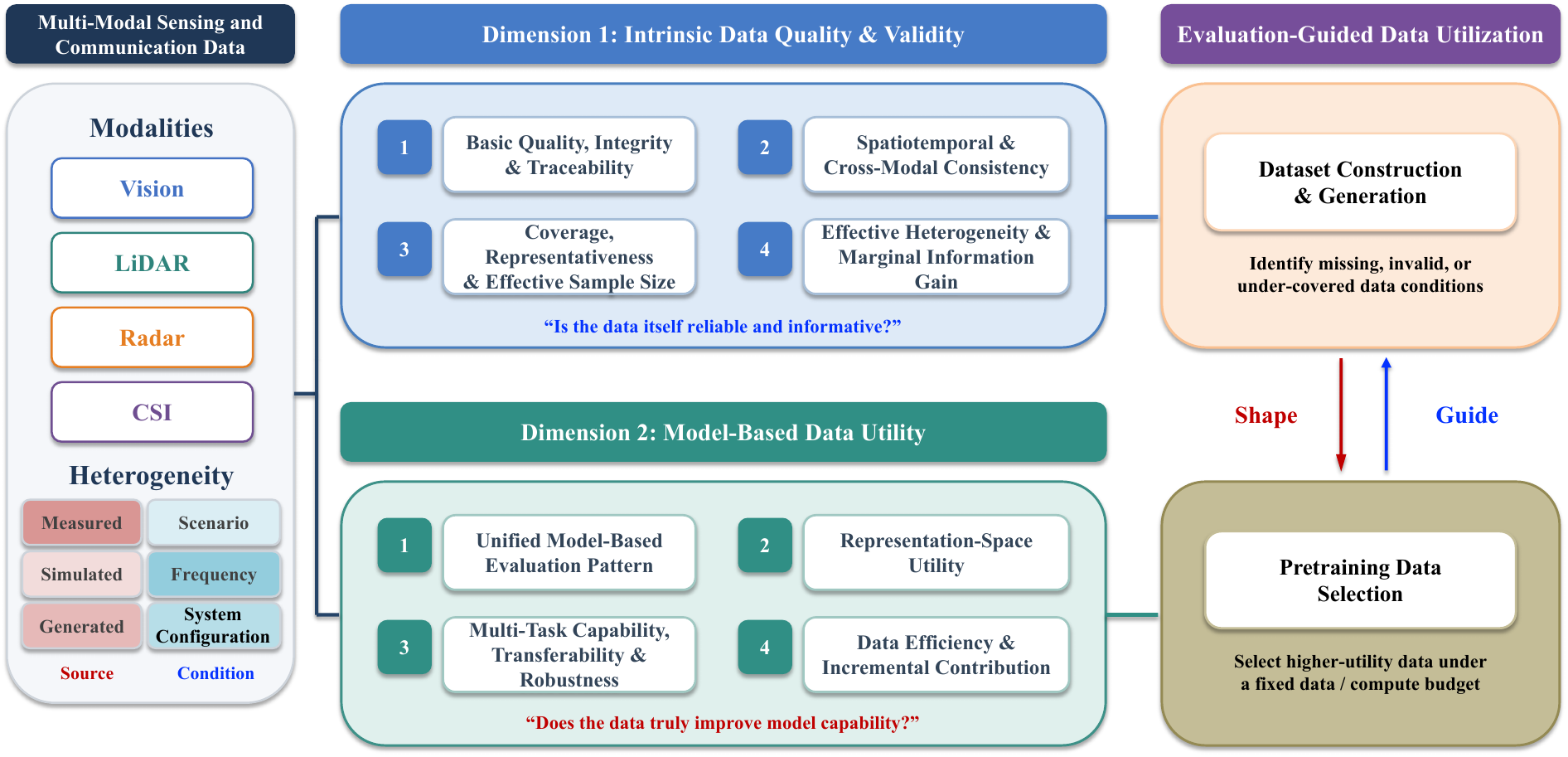}}
\caption{An illustration of the proposed multi-modal sensing and communication data evaluation framework.\label{framework}}
\end{figure*}

To comprehensively evaluate the quality of heterogeneous multi-modal sensing and communication data, we establish a data evaluation framework encompassing two dimensions: \textbf{(1) intrinsic data quality and validity} and \textbf{(2) model-based data utility}, as illustrated in Fig.~\ref{framework}.

Intrinsic evaluation assesses whether a dataset is reliable, physically reliable, and informative for pretraining without requiring full model training. The evaluation proceeds from sample integrity to physical and cross-modal consistency, then to dataset coverage and the information that new samples add. It serves two purposes: filtering defective data and identifying contributions that candidate data may make to an existing collection.

\subsection{Basic Data Quality, Integrity, and Traceability}

Quality checks should identify missing frames, dropped packets, absent modalities, outliers, severe noise, and incomplete metadata. Valid-sample ratio, signal-to-noise ratio (SNR), missing-data rate, and physical fidelity enable cross-dataset comparisons.

Traceability requires consistent metadata on origin and acquisition conditions. Measurement data should record sensing and communication configurations and calibration. Simulation data should record simulation parameter settings. Generation data should record model architectures, hyperparameters, and conditioning inputs. These records support filtering, diagnosis, and source-specific evaluation. They also support reproducibility and comparisons across data sources.

\subsection{Spatiotemporal and Cross-Modal Consistency}

Integrity alone does not ensure physical consistency. For radio-frequency (RF) data, checks should relate propagation delay to distance, Doppler shift to motion, arrival/departure angles to transceiver geometry, and spatial correlation to array configuration. Power-delay profiles, delay spread, Doppler/angular spectra, K factor, and spatial covariance further assess plausibility. These summaries can reveal distributional anomalies that physical consistency checks miss.

Multi-modal observations must align in time, space, and observed events. Evaluation should consider timestamp offsets, sampling-rate differences, coordinate transforms, sensor extrinsics, transceiver pose errors, and spatial-resolution mismatches. Misalignment can undermine otherwise valid modalities.

\subsection{Coverage, Representativeness, and Effective Sample Size}

Coverage should reflect deployment conditions, including scenario, frequency, bandwidth, devices and arrays, mobility, propagation, weather, time, and modality combinations. Comparing dataset and deployment distributions identifies underrepresented and long-tail conditions.

Nominal sample counts may be inflated by duplicates, near duplicates, and correlated observations. Condition coverage and effective sample size better describe independent dataset breadth. This matters especially when measurements cluster around a few common operating conditions.

\subsection{Effective Heterogeneity and Marginal Information Gain}

Useful heterogeneity adds distinct, task-relevant information without compromising quality or physical validity. Differences can be characterized by scenarios, system parameters, and channel statistics. A frozen sensing–communication representation model can examine feature-space coverage, clusters, and distributional distances without retraining for each dataset.

Marginal information gain measures what candidate data add beyond existing samples: new condition combinations, sparse or long-tail coverage, and nonredundant observations. This connects intrinsic validity to decisions about data collection and retention for downstream model development.

\section{Dimension 2: Model-Based Data Utility}

Intrinsic evaluation identifies valid and potentially informative data. Model-based evaluation determines whether such data improve learned capabilities through controlled experiments on representation, transfer, robustness, and marginal data value. Its central requirement is to attribute observed model gains to differences in data rather than training conditions.

\subsection{Unified Model-Based Evaluation Pattern}

We adopt a two-stage evaluation combining foundation-model fine-tuning and full pretraining. Fine-tuning offers a low-cost utility estimate: starting from identical pretrained weights, each candidate dataset adapts the same architecture with fixed training strategy, steps, and compute. Held-out performance is compared on common validation sets across repeated runs, especially when differences between datasets are small. However, fine-tuning depends on prior model knowledge and cannot establish utility for training from scratch.

The second stage validates whether these rankings predict pretraining gains. Models are pretrained with identical architectures, parameter counts, objectives, sample budgets, optimizers, training steps, and compute. Controlled comparisons report convergence and downstream task results together, since lower pretraining loss does not necessarily translate into better transferable capability.

\subsection{Representation-Space Utility}

Frozen foundation-model embeddings enable comparisons of representation distance, overlap, coverage, and within- and between-group dispersion relative to existing data and deployment conditions. Stratifying by scenario, device, frequency, and source helps distinguish new physical information from acquisition differences. Valuable additions populate sparse, task-relevant regions; near duplicates suggest redundancy, while isolated samples may reflect rare propagation, corruption, or artifacts. Physical metadata and intrinsic validity checks help identify the cause without confusing unusual conditions with defective records.

For multi-modal data, representations from aligned observations should agree more than deliberately mismatched temporal or spatial pairs. Poor separation may indicate synchronization or calibration errors, or weak encoding of shared physical state. Comparing embeddings before and after training tests whether apparent novelty becomes learned information.

\subsection{Multi-Task Capability, Transferability, and Robustness}

Evaluate utility across communication, sensing, localization, channel prediction, beam selection, environment understanding, and embodied tasks, using identical heads and adaptation budgets. Report task-specific metrics and balanced aggregates to expose tradeoffs. Generalization should be tested on held-out scenes, frequency bands, devices, and modality combinations rather than only familiar conditions.

Robustness tests should vary SNR, hardware impairment, synchronization and calibration errors, occlusion, packet loss, anomalous inputs, and modality availability, with matched perturbation severity. Evaluate both joint and single-modality inference to reflect incomplete deployment observations. Condition-specific metrics distinguish genuine robustness improvements from higher aggregate scores, including whether gains persist under realistic quality fluctuations and missing modalities.

\subsection{Data Efficiency and Incremental Data Contribution}

Data scaling laws provide a systematic framework for characterizing how model performance evolves with pretraining data volume, revealing initial gains, diminishing returns, and potential saturation. They may change with scale as diversity and condition coverage increase. Report utility per effective sample, collection cost, and training compute under common model, optimization, and evaluation protocols, especially when comparing measured, simulated, and generated data. Effective sample counts should explicitly account for near-duplicate and correlated observations. For real-simulation mixtures, vary mixing ratios to test whether simulated samples complement measured observations or create distribution gaps and diminishing marginal returns.

Measure incremental value by adding scenarios, propagation conditions, modalities, or sources to a fixed corpus. Compare against equal-size random additions, repeated records, single-condition expansions, and mismatched multi-modal pairs. For hybrid datasets, also compare measured-only, simulated-only, and mixed additions under equal budgets, accounting for simulation fidelity and domain discrepancy. Gains beyond these controls indicate nonredundant information rather than volume alone and test whether intrinsic coverage, novelty, and cross-modal consistency predict downstream utility and scaling.



\section{Case Studies: Validation of the Data Evaluation Framework}
\label{sec:case_studies}

\begin{figure*}[!t]
    \centering
    \includegraphics[width=0.95\textwidth]{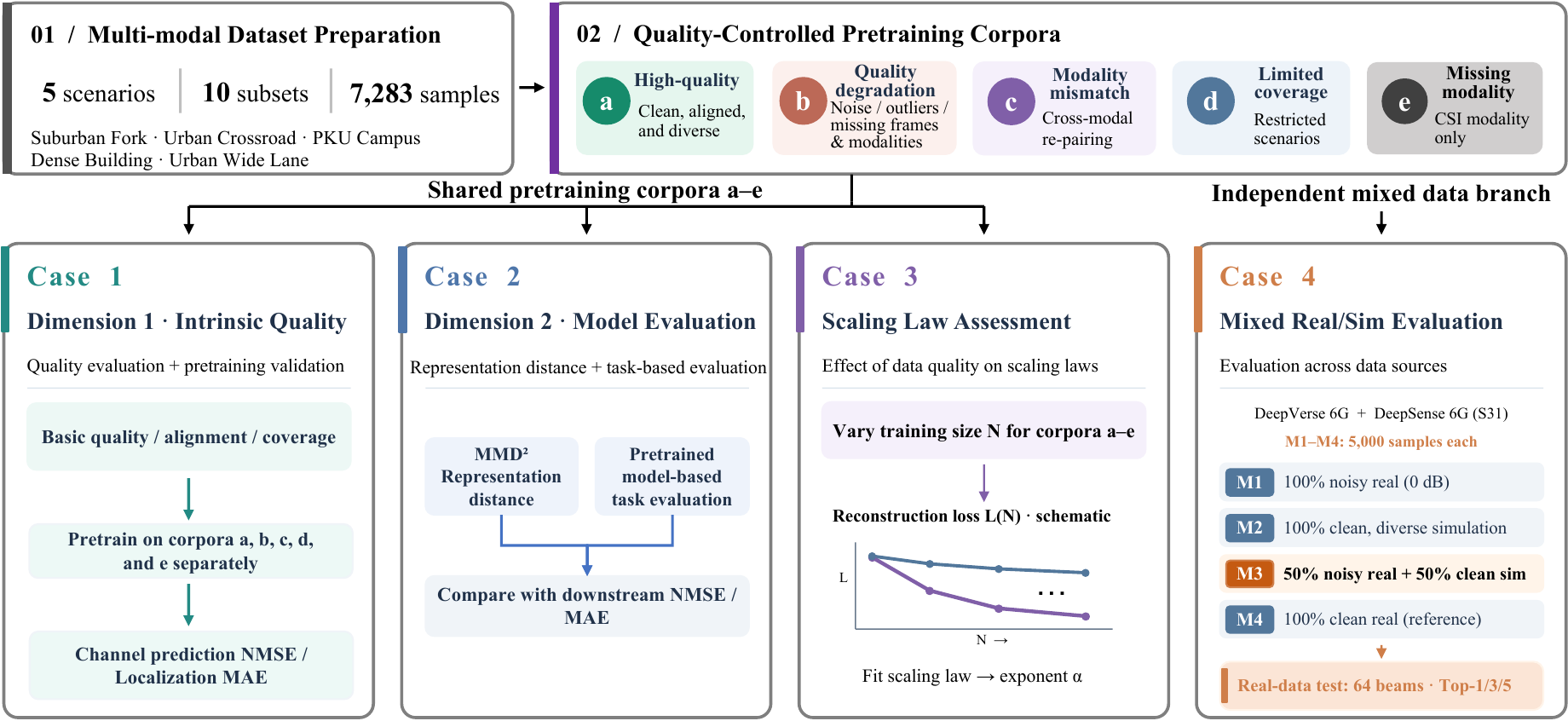}
    \caption{Overview of the experimental workflow and details of the case studies.}
    \label{cases}
\end{figure*}

\begin{figure}[!t]
    \centering
    \includegraphics[width=0.5\textwidth]{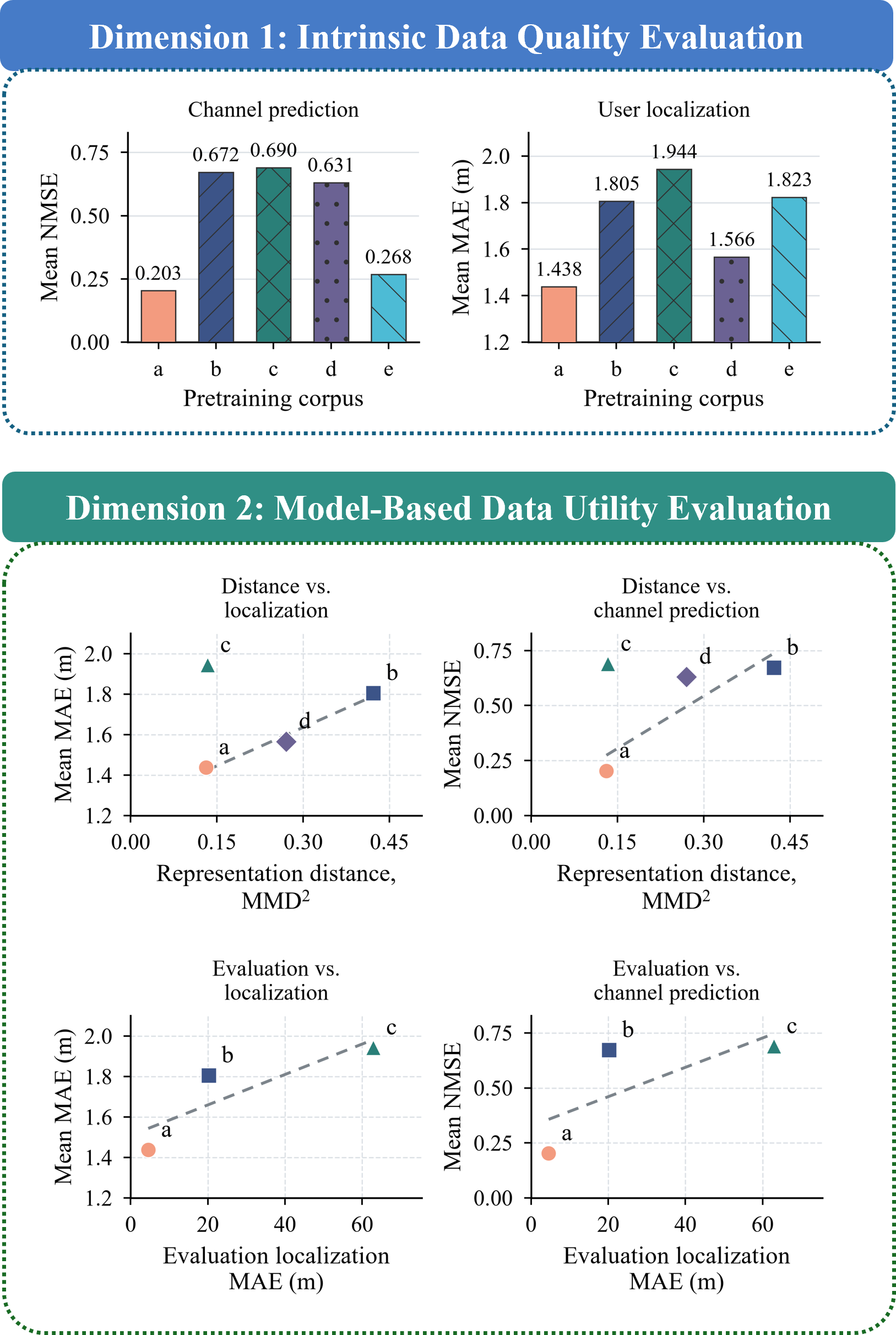}
    \caption{Evaluation of the two data-quality dimensions.}
    \label{fig:case_dimensions}
\end{figure}

To clearly illustrate the experimental workflow and key experimental details of the case studies, Fig.~\ref{cases} provides a comprehensive overview of the dataset preparation and the subsequent evaluation case studies. To illustrate how the two evaluation dimensions inform pretraining-data evaluation, we present four complementary controlled analyses, summarized in Figs.~\ref{fig:case_dimensions}--\ref{fig:case_beam64}. These figures examine the effects of intrinsic quality and modality availability on downstream performance, the usefulness and limitations of model-based evaluation indicators, data-quality-dependent scaling behavior, and the relative value of measured, simulated, and mixed training data. Together, these examples connect the proposed evaluation criteria with practical data-selection decisions.

\subsection{Experimental Setup}
\label{subsec:case_setup}

\subsubsection{Dataset Preparation}
For the intrinsic-quality and model-utility analyses, the reference corpus contains 7,283 multi-modal samples from ten subsets spanning five scenarios: 817 samples in Suburban Fork, 1,280 samples in Urban Crossroad, 2,000 samples in PKU Campus, 2,000 samples in Dense Building, and 1,186 samples in Urban Wide Lane.

We consider five pretraining corpora, denoted a--e, to isolate different aspects of data quality and modality availability. \emph{Corpus a} is the high-quality baseline, preserving observation integrity, correct cross-modal correspondence, and coverage of all five scenarios. \emph{Corpus b} introduces integrity defects, including Gaussian noise at 10~dB SNR in CSI and radar, impulsive outliers in 1\% of their elements, jointly missing CSI and radar observations for 15\% of samples, and independent radar and map modality dropout with a 20\% probability. \emph{Corpus c} disrupts cross-modal alignment by pairing observations from different samples within each subset while keeping individual modality measurements unchanged. \emph{Corpus d} uses only PKU Campus data (PF1--PF9), thereby representing reduced scenario coverage. \emph{Corpus e} is a single-modality baseline that retains only CSI data, omitting the other modalities. These configurations allow the consequences of integrity, alignment, and coverage deficiencies, as well as restricted modality availability, to be examined separately.

For the measured--simulated comparison, we use DeepVerse 6G and DeepSense 6G (Scenario~31) to construct four training-data configurations with an identical budget of 5,000 samples each. M1 consists entirely of noisy measured data at 0 dB SNR, whereas M2 consists entirely of noise-free, distribution-diverse simulated data. M3 contains an equal mixture of the two sources, with 2,500 noisy measured samples and 2,500 clean simulated samples. M4 consists of 5,000 measured samples without added noise and serves as a clean-measurement reference. All four configurations are evaluated using the same measured test data.

\subsubsection{Model Construction and Evaluation Metrics}
For Dimension~2, we use the pretrained WiFo-MiSAC \cite{liu2026wifomisac} as an evaluator. Multi-modal features extracted from this model are compared with a shared reference representation set using squared maximum mean discrepancy (MMD$^2$). We also fine-tune the evaluator on candidate data and assess localization performance using mean absolute error (MAE). This task-based evaluation is complementary to representation distance: a small distributional discrepancy does not by itself establish accurate cross-modal correspondence or high downstream utility. The same general approach can accommodate other downstream tasks and their respective performance metrics.

To obtain an independent measure of actual pretraining utility, we train a separate masked-autoencoder model on each of corpora a--e, using only CSI for corpus e. This backbone differs from the WiFo-MiSAC evaluator, reducing dependence on a single model architecture when assessing the value of the data. We evaluate the resulting models on channel prediction, measured by normalized mean squared error (NMSE), and user localization, measured by MAE in meters. We additionally compare masked-reconstruction losses across training-data budgets. In the measured--simulated study, beam prediction is evaluated using Top-1, Top-3, and Top-5 accuracy under the 64-beam setting.

\subsection{Simulation Results}
\label{subsec:case_results}

\subsubsection{Intrinsic Data Quality and Downstream Pretraining Performance}
\label{subsubsec:case_downstream}

As shown in Fig.~\ref{fig:case_dimensions} (Dimension~1), the high-quality corpus a produces the lowest channel-prediction NMSE (0.203) and user-localization MAE (1.438~m). The corresponding NMSE values for corpora b, c, d, and e are 0.672, 0.690, 0.631, and 0.268, while their localization MAEs are 1.805, 1.944, 1.566, and 1.823~m, respectively. Corpus a performs best on both tasks, whereas corpus c performs worst. Notably, the CSI-only corpus e achieves the second-lowest channel-prediction NMSE but a relatively high localization MAE, indicating task-dependent effects of restricted modality availability.

The results indicate that reliable individual observations, correct cross-modal correspondence, sufficient scenario coverage, and modality availability each contribute to the practical value of pretraining data. In particular, increasing the number of samples cannot necessarily compensate for defective alignment or corrupted measurements. Intrinsic evaluation should therefore consider multiple complementary quality properties rather than relying only on dataset size.

\begin{figure}[!t]
    \centering
    \includegraphics[width=0.5\textwidth]{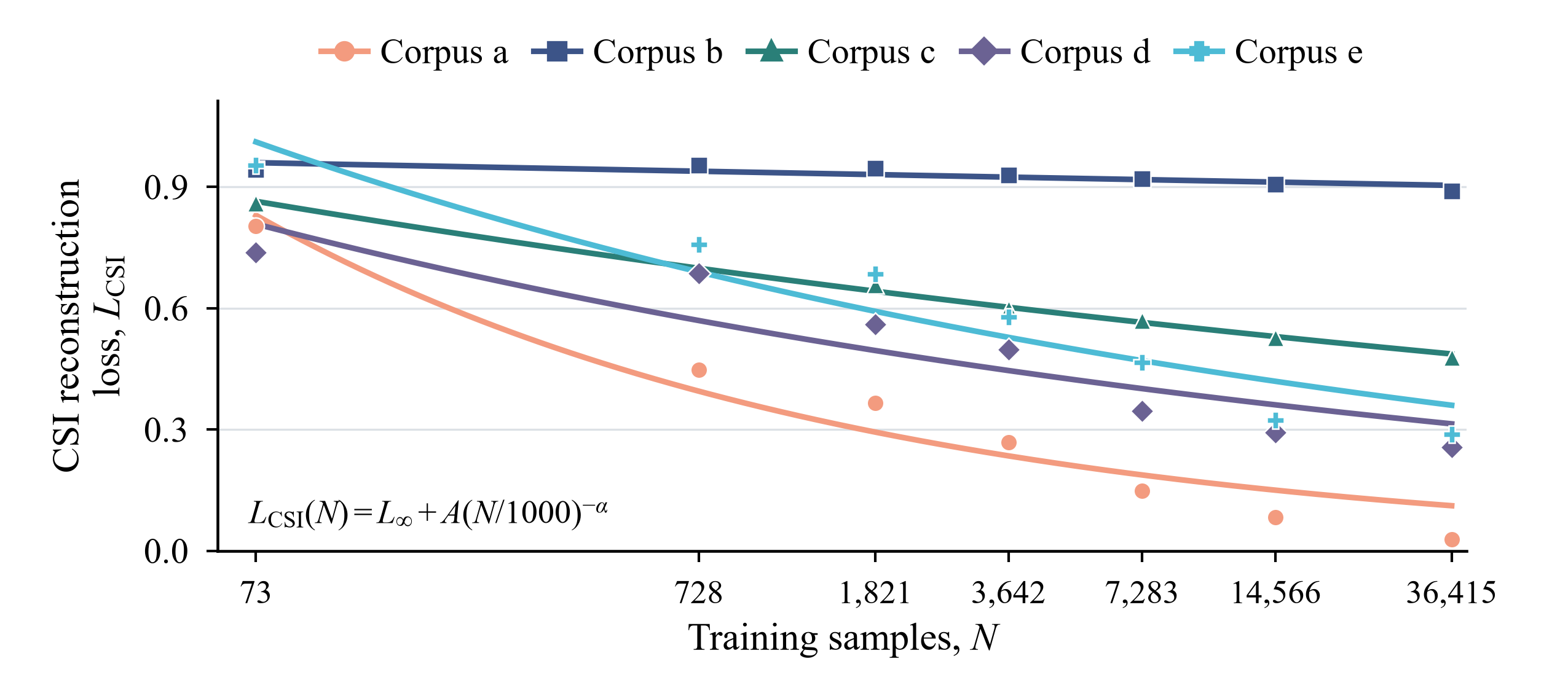}
    \caption{Data-quality-dependent scaling of CSI reconstruction loss with pretraining sample size for corpora a--e. Curves show fitted scaling-law trends.}
    \label{fig:case_scaling}
\end{figure}

\subsubsection{Validity of Model-Based Data Evaluation}
\label{subsubsec:case_validity}

Fig.~\ref{fig:case_dimensions} (Dimension~2) compares downstream performance with two model-based evaluation indicators: representation distance and the evaluator's localization error. The MMD$^2$ values for corpora a, b, c, and d are approximately 0.131, 0.422, 0.133, and 0.270, respectively. Although representation distance distinguishes some quality differences, the misaligned corpus c lies almost as close to the reference representation distribution as corpus a, despite yielding the largest downstream errors. Thus, a distribution-level similarity measure can overlook incorrect correspondence between modalities.

Task-based evaluation provides a useful complementary signal. For corpora a, b, and c, the evaluator's localization MAEs are approximately 4.55, 20.23, and 62.89~m, respectively, consistent with the deterioration observed in the independently pretrained models. Since corpus d uses a different evaluation set, its evaluator error is not used for direct ranking against a--c. This comparison illustrates why representation-based screening and task-based validation are best considered together, particularly when multi-modal alignment is a potential source of error.

\subsubsection{Data Quality and Scaling Behavior}
\label{subsubsec:case_scaling}

Fig.~\ref{fig:case_scaling} examines masked-reconstruction loss as the pretraining-data budget increases from 73 to 36,415 samples. The empirical scaling trends differ substantially across the pretraining corpora. The high-quality corpus a benefits strongly from additional samples, while the integrity-degraded corpus b improves only marginally; corpora c and d exhibit intermediate behavior. The CSI-only corpus e also benefits from scaling, with reconstruction loss decreasing to approximately 0.29 at the largest budget. At this budget, the observed losses for a and b are approximately 0.035 and 0.592, respectively.

An important implication is that the benefit of increasing data volume depends on the quality of the additional observations. The advantage of corpus a becomes more pronounced at larger training budgets, even though it does not achieve the lowest reconstruction loss at the smallest budget. These results concern pretraining reconstruction loss and should not, without further downstream testing, be interpreted as a universal guarantee of improved generalization. In particular, the CSI-only loss of corpus e should be interpreted with its different modality coverage in mind.

\begin{figure}[!t]
    \centering
    \includegraphics[width=0.5\textwidth]{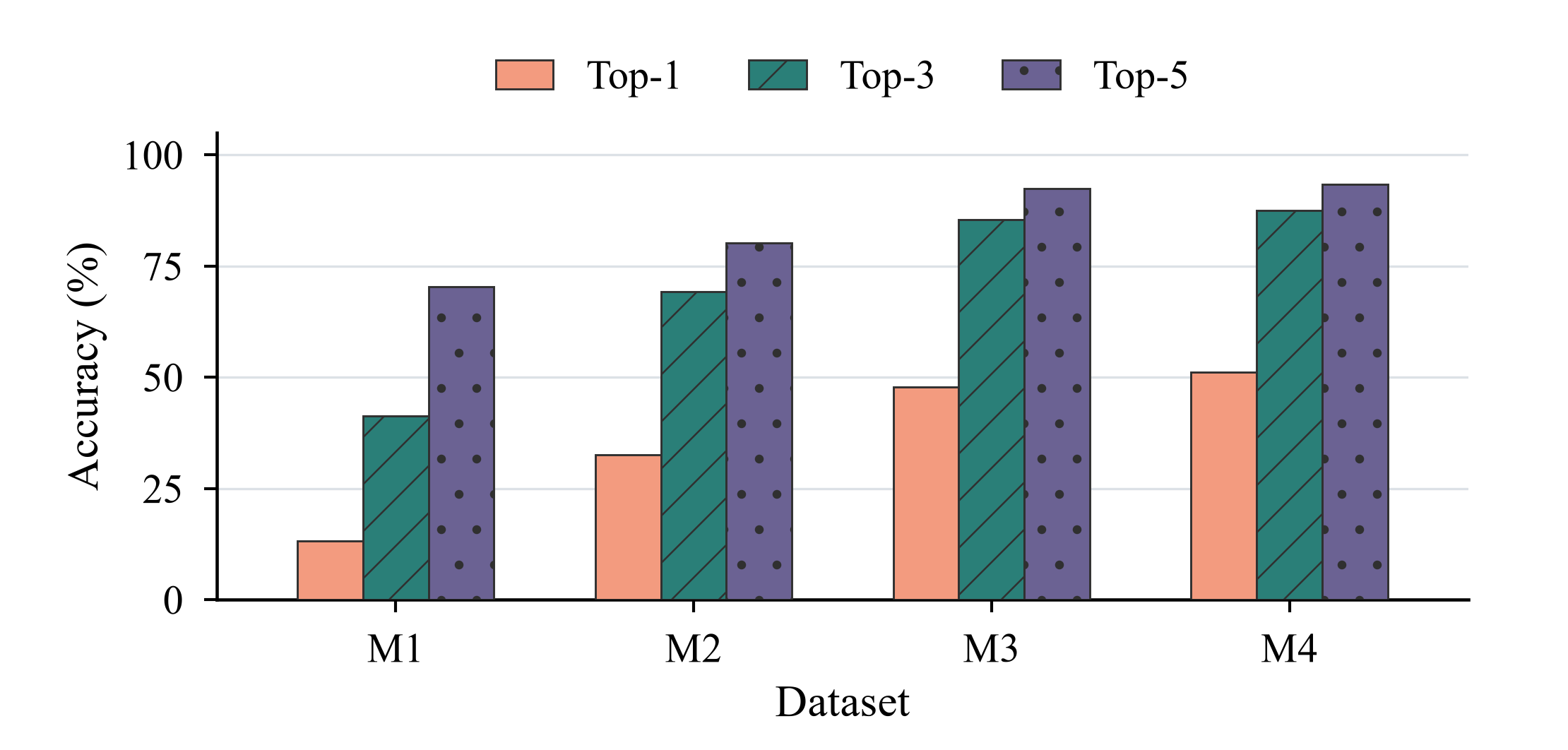}
    \caption{Top-1, Top-3, and Top-5 accuracy for 64-beam prediction on the measured test set.}
    \label{fig:case_beam64}
\end{figure}

\subsubsection{Measured-Simulated Mixtures for Beam Prediction}
\label{subsubsec:case_beams}

Fig.~\ref{fig:case_beam64} compares the four data-source configurations on the measured 64-beam prediction test set. Top-1 accuracy rises from 13.2\% with noisy measured-only training (M1) to 32.6\% with clean, diverse simulated data (M2). Under these controlled conditions, simulated observations of higher integrity and diversity are therefore more useful than noisy measurements, even when performance is tested on measured data.

Combining the two sources yields further gains. M3 reaches a Top-1 accuracy of 47.8\% for 64-beam prediction, approaching the clean measured-data reference M4 at 51.2\%. Top-3 and Top-5 accuracies exhibit similar improvements. The comparison suggests that source provenance alone is an insufficient indicator of data value: noise, validity, distributional diversity, and source complementarity can be more consequential. Carefully designed measured-simulated mixtures can mitigate limitations of noisy measured data while retaining useful diversity from simulation.

Overall, these four analyses illustrate the complementary roles of intrinsic and model-based evaluation. Integrity, alignment, coverage, and modality-availability checks characterize defects and omissions before training; representation- and task-based indicators help reveal their likely model-level consequences. Scaling and source-mixing experiments further highlight that pretraining data should be evaluated not only by quantity and origin, but also by quality, diversity, and demonstrated downstream utility.

\section{Discussions and Open Issues}

\subsection{Quantifying Data Quality}

The quality of multi-modal sensing and communication data is determined by multiple dimensions. However, different dimensions, as well as individual factors within each dimension, may contribute unequally to data quality. Their complex interactions and potential trade-offs make it challenging to aggregate them into a unified score using predefined weights. Future research should  focus on developing quantitative metrics to characterize the contributions of individual factors and systematically evaluating their relative importance. Establishing interpretable weighting mechanisms and evaluation frameworks could enable more accurate and reliable quantification of data quality.

\subsection{High-Quality Multi-Modal Measurement Platforms}

Obtaining reliable ground-truth data remains a fundamental challenge for both RF channel and multi-modal sensing. Future research should focus on developing high-fidelity multi-modal measurement platforms that capture real-world RF channel responses alongside synchronized sensing data in diverse physical environments. Such platforms could provide authentic measurement benchmarks, enabling systematic evaluation of data fidelity, cross-modal consistency, and physical coherence, thereby establishing a reliable foundation for multi-modal data evaluation.

\subsection{Low-Cost Prediction of Model Utility}

Training multi-modal foundation models requires substantial computational resources\cite{wcm-train}. Consequently, evaluating data quality through full-scale model pretraining and downstream performance testing is computationally expensive, limiting its scalability for large-scale data selection. Future research should explore lightweight proxy models and computationally efficient methods for estimating dataset utility across diverse tasks and model architectures. Establishing a reliable mapping from intrinsic data properties to model utility could enable rapid quality evaluation and efficient pretraining data selection without costly model-based experiments.

\section{Conclusions}
In this paper, we have proposed a comprehensive data evaluation framework for multi-modal sensing and communication foundation models. We have first reviewed existing data evaluation methods in NLP, CV and multi-modal perception, and wireless communications. Nevertheless, existing methods lack unified evaluation of heterogeneous multi-modal data, thereby hindering multi-modal pretraining dataset construction and selection. To fill this gap, we have established a comprehensive evaluation framework from two complementary dimensions, including intrinsic data quality and validity, and model-based data utility. Furthermore, we have conducted four case studies to validate the proposed framework. We have separately validated the effectiveness of the two evaluation dimensions, further investigated the impact of data quality on pretraining scaling laws, and finally demonstrated that appropriately integrating measurement and simulation data can compensate for the limitations of low-quality measured data. Finally, we have discussed promising future research directions, including the transition from static data quality to dynamic data value and the development of low-cost and reliable methods for predicting model utility.

\newpage

 





\end{document}